\documentclass[seceq]{ptptex}
\usepackage{graphicx} 
\usepackage{latexsym}
\notypesetlogo
\begin{document}
%\begin{flushright} 
%RUP- \\
%
%\end{flushright}
\vspace{10mm}
%\begin{document}
\begin{center}
\Large{ \bf Schwinger-Dyson equation in the complex plane }

\vspace{5mm}

\Large{\bf $-$ Two simple models $-$}

\vspace{15mm}
   
\large{ Hidekazu {\sc Tanaka} \footnote{E-mail:tanakah@rikkyo.ac.jp} and Shuji {\sc Sasagawa} \\
Rikkyo University, Tokyo 171-8501, Japan\\
}
 \end{center}

\begin{center}

\vspace{25mm}

{\Large ABSTRACT}
 \end{center}
        
\vspace{10mm}

Effective mass and energy are investigated using the Schwinger-Dyson equation (SDE) in the complex plane.
As simple examples, we solve the SDE for the (1+1)-dimensional model and the strongly coupled quantum electrodynamics (QED). We also study some properties of the effective mass and energy in the complex plane.

%\
%\markboth{H. {\sc Tanaka}   }{}
\def\proj{{\bf P}}
\def\slsh#1{{#1}{\kern-6pt}/{\kern1pt}}
%\title{ }

%\vspace{15mm}

%\author{Hidekazu {\sc Tanaka}}
%\inst{ Rikkyo University, \\
%           Nishi-ikebukuro, Toshima-ku Tokyo, Japan, 171 \\
%     }
%\recdate{ }
%  \begin{center}
%\maketitle
%\vspace{25mm}

%  {\Large ABSTRACT}
%  \end{center}
      
%\vspace{10mm}

% \maketitle

%  \newpage

\newpage

\section{Introduction}

The Schwinger-Dyson equation (SDE) [1,2] is one of the useful tools for evaluating non-perturbative phenomena.
In many studies, the SDE is defined in Euclidean space.
However, various extensions may be useful in computing the SDE, such as application to Minkowski space and extension to the complex momentum space.

One of the difficulties in calculation in Minkowski space is the presence of poles in the propagator. This requires knowledge of the precise pole positions of the propagator in the self-energy calculation. 
To avoid this, the Wick rotation from the real axis to the imaginary axis is often used in the momentum integration. However, the Wick rotation requires the location of the poles to be known in advance, but the value of an effective mass in  non-perturbative region is non-trivial in calculation with the SDE. \footnote{ A formal approach has been done in Refs.[3,4], in which the gluon propagator with complex singularities in Minkowski space is reconstructed starting from the Euclidean propagator by the analytic continuation.}

In order to study the properties of the propagators in non-perturbative region, the squared  four-dimensional  momentum has been extended to the complex value for the electron propagator in quantum electrodynamics (QED) [5,6] and the gluon propagator in QCD [7-9].

In this paper, we investigate a method, in which the mass and energy of the SDE are extended to the complex plane, and the propagator in the integrand of the self-energy is integrated with two different integral paths in the complex energy plane.
 Then we examine the properties of the solutions obtained by the SDE in the complex plane. 
 
 We explain our formulation and analysis method using two simple models.
 First, in order to illustrate our method, we examine a simple SDE for the effective mass which does not depend on the energy and momentum.
This model in two-dimensions can be solved analytically and we compare these results with the numerically determined solutions obtained by the SDE.
 As a second model, we investigate the electron mass function in the strongly coupled QED with the instantaneous exchange approximation (IEA) [10,11], which is a slightly extended model of the first example.

Our paper is organized as follows:
In Section 2, we formulate the SDE for the effective mass with a complex mass and energy for the first model. We discuss analytical solutions for the effective mass and energy in the complex plane with finite cutoff values of the momentum and we numerically calculate the effective mass and energy using the SDE. 
In Section 3, using the method presented in section 2, we investigate the SDE for the electron in the strongly coupled QED with the IEA.
Section 4 is devoted to a summary and some comments. Explicit expressions of the complex mass and energy implemented in calculations are given in Appendix A. In Appendix B, we present a method to obtain the mass function in Minkowski space from the mass function obtained by the SDE in Euclidean space.

%\newpage

\section{The SDE for a simple model in the complex plane}

In this section, we  investigate  a simple SDE for the effective mass $M$  in order to illustrate our method.  \footnote{This example is a simplified model of the SDE for the effective mass of models with four-fermion interactions in (1+1)-dimensions, such as the Gross-Neveu model [12]. } The SDE for the effective mass is given as
\begin{eqnarray}
 M=i\lambda \int d^2Q {M \over Q^2-M^2+i\varepsilon}
= i\lambda \int dq_0dq {M \over q_0^2-q^2-M^2+i\varepsilon}
\end{eqnarray}
with a momentum $Q=(q_0,q)$.  Here, $\lambda$ is a coupling constant.

The propagator in the integrand has poles, which satisfy $Q^2-M^2+i\varepsilon = q_0^2-q^2-M^2+i\varepsilon =0$.

In order to evaluate the effective mass in the complex plain, we extend $q_0$ to a complex value $z$ and the effective mass $M$ is also extended to a complex value. Explicitly, they are written as $z \equiv z_{\rm R}+iz_{\rm I}$ and $M\equiv     M_{\rm R}+iM_{\rm I}$, respectively. Here, we write the denominator in the integrand as 
\begin{eqnarray}
z^2-q^2-M^2+i\varepsilon\equiv z^2-E^2(q)= (z-E(q))(z+E(q)),
\end{eqnarray}
where the complex energy $E(q)$ is written as 
\begin{eqnarray}
E(q)\equiv E_{\rm R}(q)+iE_{\rm I}(q).
\end{eqnarray}
Therefore, the poles are located at $z=\pm E(q)$ in the complex $z$ plane. Here, we define $ E_{\rm R}(q)> 0$. Explicit relations among the complex values are given in Appendix A.

The SDE for the effective mass in terms of the complex values is written as
$$
 M=i\lambda\int dq  \oint_C dz {M \over  z^2-q^2-M^2(q)+i\varepsilon} 
$$
\begin{eqnarray}
= i\lambda \int dq  \oint_C dz {M \over  (z- E(q))(z+ E(q))}.
\end{eqnarray}

Here, we write above equation as 
\begin{eqnarray}
 M={1 \over 2}M^{(+)}+{1 \over 2}M^{(-)},
\end{eqnarray}
where
$$M^{(\pm)} =i\lambda\int dq  \oint_C dz{1\over  z -z_{\pm} }\left[{M \over z + z_{\pm}}\right] $$
\begin{eqnarray}
 \equiv i\lambda\int dq  \oint_C dz{1\over  z -z_{\pm} }f^{(\pm)}(z,q)
 \end{eqnarray}
with $z_{\pm}=\pm E(q)$ and 
\begin{eqnarray}
f^{(\pm)}(z,q)={M \over z + z_{\pm}}.
\end{eqnarray}

In our calculation, we integrate Eq.(2$\cdot$6) around $z=z_{\pm}$ with following two integral paths. 

\vspace{5mm}

(1) Integral path including the imaginary axis

\vspace{5mm}

In this case, we separate the integral path around the poles $z_{\pm}=\pm E(q)$ to $C_1$ and $C_2$ as follows:

For the integral path around $z_+=E(q)$, we take $-i\Lambda_0 \leq z \leq i\Lambda_0$ as the path $C_1$, and the path $C_2$ is defined as clockwise rotation in right-half on the complex energy plane with $z=\Lambda_0e^{i\theta}$, where we  integrate the angle $\theta$  from $\theta=\pi/2$ to $\theta=-\pi/2 $.
  
On the other hand, for the integral path around $z_-=-E(q)$, we take $-i\Lambda_0 \leq z \leq i\Lambda_0$ as the path $C_1$, and the path $C_2$ is defined  as anticlockwise rotation in left-half on the complex energy plane with $z=\Lambda_0e^{i\theta}$, where we  integrate the angle $\theta$  from $\theta=\pi/2$ to $\theta=3\pi/2 $. 
 
Integrating over the integral path $C$ around the poles $z_{\pm}=\pm E(q)$ in the right-hand side of Eq. (2$\cdot$6), we have  
\begin{eqnarray}
M^{(\pm)}=i\lambda \int dq (\mp 2\pi i)f^{(\pm)}(z_{\pm},q)
=\pi\lambda\int dq {M \over  E(q)} \end{eqnarray}
for $\Lambda_0\rightarrow \infty$.
Therefore, the SDE for the effective mass is given by
\begin{eqnarray}
 M={1 \over 2}M^{(+)}+{1 \over 2}M^{(-)}
=\pi\lambda\int dq {M \over  E(q)}.
\end{eqnarray}

 This case corresponds to the SDE for  the  Euclidian momentum integration with the complex mass $M$, which is given as
\begin{eqnarray}
 M = \lambda\int dq  \int_{-\infty }^{\infty } dq_4 {M \over  q_4^2+q^2+M^2-i\varepsilon} 
\end{eqnarray}
with $z=iq_4$ in Eq. (2$\cdot$4).\footnote{ The contribution from the integral path $C_2$ vanishes for $\Lambda_0\rightarrow \infty$.}
\vspace{5mm}

(2) Integral path including the real axis

\vspace{5mm}

In this case, we separate the integral path around the poles $z_{\pm}=\pm E(q)$ to $C_1$ and $C_2$ as follows:

For the integral path around $z_+=E(q)$, we take $-\Lambda_0 \leq z \leq \Lambda_0$ as the path $C_1$, and if $E_{\rm I}>0$, the path $C_2$ is defined as anticlockwise rotation  in upper-half on the complex energy plane with $z=\Lambda_0e^{i\theta}$, where we  integrate the angle $\theta$  from $\theta=0$ to $\theta=\pi $. If $E_{\rm I}< 0$, we take $-\Lambda_0 \leq z \leq \Lambda_0$ as the path $C_1$, and the path $C_2$ is defined as clockwise rotation  in lower-half on the complex energy plane with $z=\Lambda_0e^{i\theta}$, where we  integrate the angle $\theta$  from $\theta=0$ to $\theta=-\pi $. 
  
For the integral path around $z_-=-E(q)$, we take $-\Lambda_0 \leq z \leq \Lambda_0$ as the path $C_1$, and if $E_{\rm I}<0$, the path $C_2$ is defined as anticlockwise rotation in upper-half on the complex energy plane with $z=\Lambda_0e^{i\theta}$, where we  integrate the angle $\theta$  from $\theta=0$ to $\theta=\pi $. If $E_{\rm I}> 0$, we take $-\Lambda_0 \leq z \leq \Lambda_0$ as the path $C_1$, and the path $C_2$ is defined as clockwise rotation  in lower-half on the complex energy plane with $z=\Lambda_0e^{i\theta}$, where we  integrate the angle $\theta$ from $\theta=0$ to $\theta=-\pi $. 

Integrating over the integral path $C$ around the poles $z_{\pm}=\pm E(q)$ in the right-hand side of Eq. (2$\cdot$6), we have  
$$M^{(\pm)} =i\lambda \int dq (\pm 2\pi i)\left({E_{\rm I}(q)\over |E_{\rm I}(q)|}\right)f^{(\pm)}(z_{\pm},q)$$
\begin{eqnarray}
=-\pi\lambda\int dq \left({E_{\rm I}(q)\over |E_{\rm I}(q)|}\right){M(q) \over  E(q)} \end{eqnarray}
for $\Lambda_0\rightarrow \infty$.

Therefore, the effective mass is given by
\begin{eqnarray}
 M={1 \over 2}M^{(+)}+{1 \over 2}M^{(-)}
=-\pi\lambda\int dq \left({E_{\rm I}(q)\over |E_{\rm I}(q)|}\right){M \over  E(q)}. \end{eqnarray}

This case corresponds to the SDE for  the  Minkowski momentum integration with the complex mass $M$, which is given as
\begin{eqnarray}
 M= i\lambda\int dq  \int_{-\infty }^{\infty } dq_0 {M \over  q_0^2-q^2-M^2+i\varepsilon}
\end{eqnarray}
 with $z=q_0$ in Eq. (2$\cdot$4).
 
\subsection{Analytical solutions}

In this model, the effective mass does not depend on the momentum, and we can find analytical solutions for the SDE. Here, we write the SDE for the two integral paths (1) and (2) as
\begin{eqnarray}
 M=\pi s\lambda\int dq {M \over E(q)}
\end{eqnarray}
with $s=1$ for the integral path (1) and $s=-E_{\rm I}(q)/| E_{\rm I}(q)|=-[(M^2)_{\rm I}-\varepsilon ]/| (M^2)_{\rm I}-\varepsilon |$ for the integral path (2), respectively. 

From Eq.(2$\cdot$14), the nontrivial solutions with $M\neq 0$ are given by solving the equation  
\begin{eqnarray}
1-\pi s\lambda\int dq {1 \over E(q)} =0.
\end{eqnarray}
Thus, the real and imaginary parts of Eq.(2$\cdot$15) satisfy 
\begin{eqnarray}
1-\pi s\lambda\int dq {E_{\rm R}(q) \over |E(q)|^2}=0 \end{eqnarray}
and 
\begin{eqnarray}
\pi s\lambda\int dq {E_{\rm I}(q) \over |E(q)|^2}=0, 
\end{eqnarray}
respectively.

Here, Eq. (2$\cdot$17) is written as
\begin{eqnarray}
\pi s\lambda\int dq {E_{\rm I}(q) \over |E(q)|^2}
=\pi s\lambda((M^2)_{\rm I}-\varepsilon)\int dq {1 \over 2E_{\rm R}(q) |E(q)|^2}=0.\end{eqnarray}

For $E_{\rm R}(q)>0$, we obtain $(M^2)_{\rm I}-\varepsilon=0$ for $s\lambda\neq 0$, which gives 
$E_{\rm I}(q)=0$.\footnote{As shown in the next subsection, $ E_{\rm I}(q)$ is determined by the asymptotic value, which is  numerically calculated by the SDE with a given initial value of the mass $M$.} 

Moreover, from 
\begin{eqnarray}
 (M^2)_{\rm I}-\varepsilon=2M_{\rm R}M_{\rm I}-\varepsilon=0,\end{eqnarray}
the imaginary part of the effective mass $ M_{\rm I}$ is given by
\begin{eqnarray}
 M_{\rm I}={\varepsilon \over 2M_{\rm R}}.\end{eqnarray}
In the calculation below, we neglect the imaginary part of the effective mass for small $\varepsilon$.
Thus, we approximate as $(M^2)_{\rm R}\simeq M_{\rm R}^2$ for simplicity.

Using $E_{\rm I}(q)=0$,  and introducing a ultraviolet cutoff $\Lambda$ and an infrared cutoff $\delta$ for the momentum $q$, we write  Eq. (2$\cdot$16) as
\begin{eqnarray}1=\pi s\lambda\int_{-\Lambda}^{\Lambda} dq {1 \over  E_{\rm R}}\Theta(|q|-\delta)
=2\pi s\lambda\int_{\delta}^{\Lambda} dq {1 \over  \sqrt{q^2+ M_{\rm R}^2}}.
\end{eqnarray}
Here, $\Theta(|q|-\delta)$ denotes the step function for restriction of the momentum $q$.

Eq. (2$\cdot$21) gives
\begin{eqnarray}
1=2\pi s\lambda \log\left|{\Lambda+\sqrt{\Lambda^2+ M_{\rm R}^2}\over \delta+\sqrt{\delta^2+ M_{\rm R}^2}}\right|,\end{eqnarray}
 which is satisfied if $s\lambda>0$.
 Therefore, for $\lambda>0$, $s=-E_{\rm I}(q)/| E_{\rm I}(q)|=1$ should be satisfied for the integral path (2). 

Defining $m_{\rm R}= M_{\rm R}/\Lambda$, ${\bar \delta}=\delta/\Lambda$ 
 and 
\begin{eqnarray} {1+\sqrt{1+m_{\rm R}^2}\over {\bar \delta}+\sqrt{{\bar \delta}^2+m^2_{\rm R}}}=e^{1/(2\pi s\lambda)}\equiv \zeta \end{eqnarray}
  from Eq. (2$\cdot$22),  Eq. (2$\cdot$23) is written as 
\begin{eqnarray}m_{\rm R}^2(Am_{\rm R}^2-B)=0 \end{eqnarray}
 with $ A=(1-\zeta ^2)^2$ and $ B=4\zeta (1-{\bar \delta}\zeta)( \zeta -{\bar \delta})$.

 The solution for $m_{\rm R}^2\neq 0$ is given as
\begin{eqnarray} m_{\rm R}^2={B \over A}
={4\zeta (1-{\bar \delta}\zeta)( \zeta -{\bar \delta}) \over (1-\zeta ^2)^2}.\end{eqnarray}
 
 For $s\lambda > 0$, $ \zeta -{\bar \delta}> 0$ is satisfied.
Moreover, $ m_{\rm R}^2>0$ demands  $1-{\bar \delta}\zeta > 0$, which gives 
\begin{eqnarray} \zeta =e^{1/(2\pi s\lambda)}< {1 \over {\bar \delta}}={\Lambda \over \delta}. \end{eqnarray}
 Eq. (2$\cdot$26) restricts the coupling constant $\lambda$ as 
  \begin{eqnarray} \lambda > {1 \over 2\pi\log {\Lambda \over \delta}} \equiv \lambda_c\end{eqnarray}
 with $s=1$.

For above restriction of $\lambda$, the real part of the effective mass is given as
  \begin{eqnarray}
m_{\rm R}={M_{\rm R} \over \Lambda} = \pm \sqrt{{4\zeta (1-{\bar \delta}\zeta)( \zeta -{\bar \delta}) \over (1-\zeta ^2)^2}}.
  \end{eqnarray}
%\newpage
\subsection{Numerical solutions}

In this subsection,  we numerically calculate the effective mass and energy using the SDE  for the two integral paths (1) and (2).
The SDE is given in Eq. (2$\cdot$14). In numerical calculation, we write the SDE for the real and imaginary parts of the  effective  mass as 
\begin{eqnarray}M_{\rm R}=2\pi s\lambda\int_{\delta}^{\Lambda} dq {[M(E(q))^*]_{\rm R} \over |E(q)|^2}=2\pi s\lambda\int_{\delta}^{\Lambda} dq { M_{\rm R}E_{\rm R}(q)+M_{\rm I}E_{\rm I}(q) \over |E(q)|^2}     \end{eqnarray} and
\begin{eqnarray}M_{\rm I}=2\pi s\lambda\int_{\delta}^{\Lambda} dq {[M(E(q))^*]_{\rm I} \over |E(q)|^2}=2\pi s\lambda\int_{\delta}^{\Lambda} dq { M_{\rm I}E_{\rm R}(q)-M_{\rm R}E_{\rm I}(q) \over |E(q)|^2},\end{eqnarray}
respectively with $|E(q)|^2= E_{\rm R}^2(q)+ E_{\rm I}^2(q)$.
%\newpage
We solve the SDE by iteration method from some initial input values for the real and imaginary parts of the effective  mass denoted by $M_{\rm R(0)} $ and $M_{\rm I(0)} $.

 For the integral path (1), we can start from any initial input values of the effective mass to solve the SDE, since $s$ does not depend on the mass.
 However, for the integral path (2), the SDE has non-trivial solutions only for $s=-E_{\rm I}(q)/| E_{\rm I}(q)|=-[(M^2)_{\rm I}-\varepsilon ]/| (M^2)_{\rm I}-\varepsilon |=1$.
 Since $(M^2)_{\rm I}=2M_{\rm R}M_{\rm I}$, we set  $M_{\rm R(0)} $ and $M_{\rm I(0)} $, 
 which satisfy $(M^2)_{\rm I(0)}=2M_{\rm R(0)} M_{\rm I(0)} <0$.
\begin{figure}
\centerline{\includegraphics[width=8cm]{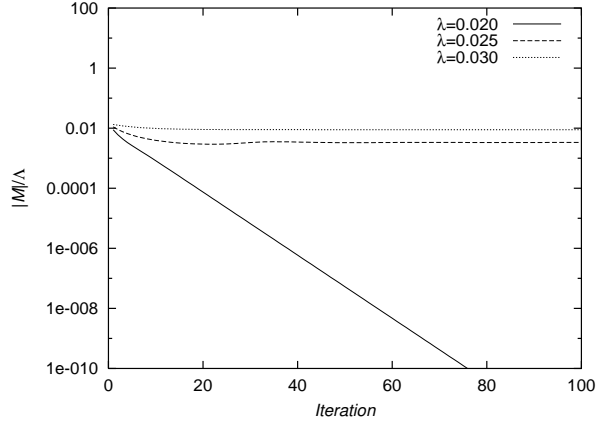}}
\caption{The convergence behaviors of $|M|/\Lambda$ for $\lambda=0.020,0.025,0.030$ with $ M_{\rm R(0)}=- M_{\rm I(0)}=0.01\Lambda$. The horizontal axis denotes the number of iterations.}
%\label{Fig.1}
\end{figure}
\hspace{15cm}

In Fig.1, we present the convergence behaviors of $|M|/\Lambda=\sqrt{M_{\rm R}^2+M_{\rm I}^2}/\Lambda$ near the critical coupling constant $\lambda_c$ denoted in Eq. (2$\cdot$27) with $\delta/\Lambda=10^{-3}$, which gives   $\lambda_c \simeq  0.023$. 
Here, we set $ M_{\rm R(0)}=- M_{\rm I(0)}=0.01\Lambda$.\footnote{We set $\varepsilon=10^{-5}\Lambda^2$.}

From Fig.1, we can conclude that $\lambda_c$ locates between $\lambda=0.020$ and $\lambda=0.025$. 
 \begin{figure}
\centerline{\includegraphics[width=8cm]{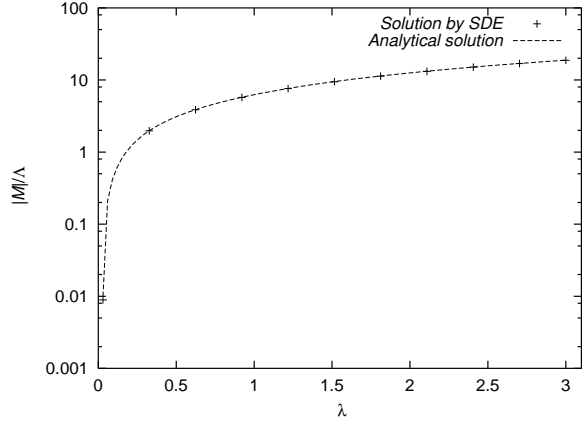}}
\caption{The $\lambda$ dependence of $|M|/\Lambda$ for $0.03 \leq \lambda \leq 3$ with $ M_{\rm R(0)}=- M_{\rm I(0)}=0.01\Lambda$.  The dotted curve denotes the calculated result by the analytical solution divided by $\Lambda$.}
%\label{Fig.1}
\end{figure}
\newpage
In Fig.2, we present the $\lambda$ dependence of the absolute value of the effective mass $|M|/\Lambda$ for $0.03 \leq \lambda \leq 3$ with $ M_{\rm R(0)}=- M_{\rm I(0)}=0.01\Lambda$. The dotted curve denotes the calculated result using Eqs. (2$\cdot$20) and (2$\cdot$28).
%\newpage
\hspace{35cm}

In the following calculations, we set four initial input values for the  effective   mass as $ M_{\rm R(0)}= \pm 0.01\Lambda$ and $M_{\rm I(0)}=\pm 0.01\Lambda$, respectively, for $\lambda=0.3$.
%\vspace{5cm}
\begin{figure}
\centerline{\includegraphics[width=8cm]{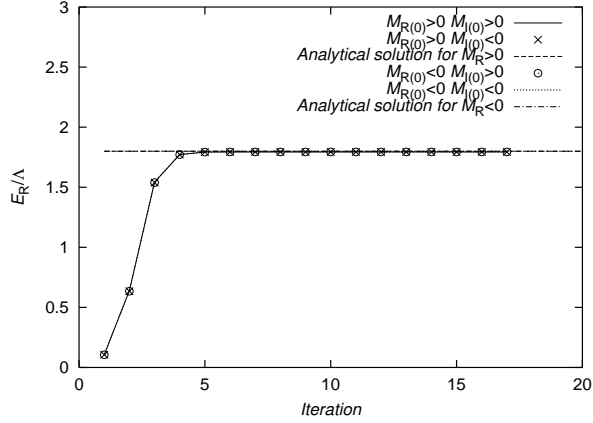}}
\caption{ The convergence  behaviors  of $ E_{\rm R}(q)/\Lambda$ with $q=0$. The straight lines denote the energy divided by $\Lambda$ calculated using the analytical solutions of the  effective  mass. The horizontal axis denotes the number of iterations.}
%\label{Fig.1}
\end{figure}
%\hspace{28cm}
%\newpage
\begin{figure}
\centerline{\includegraphics[width=8cm]{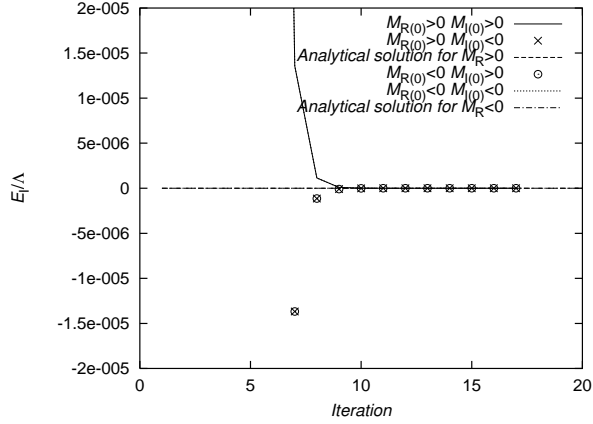}}
\caption{ The convergence  behaviors  of $E_{\rm I}(q)/\Lambda$ with $q=0$. The straight lines denote the analytical solutions of the energy, which is $E_{\rm I}(q) /\Lambda =0$. The horizontal axis denotes the number of iterations.}
%\label{Fig.1}
\end{figure} 
\hspace{13cm}

In Figs.3 and 4, we show the convergence behaviors of the real and imaginary parts of the energy with the momentum $q=0$, respectively. The straight lines denote the energy calculated using the analytical solutions of the  effective  mass given in Eqs.(2$\cdot$20) and (2$\cdot$28).(See Appendix A.)

Since the real part of the energy is defined to be positive, the numerical results do not depend on the  signs  of initial input values of the  effective  mass.
The imaginary part of the energy converges as $E_{\rm I}\rightarrow 0$, in which the convergence  behavior depends  on the  respective signs of $(M^2)_{\rm I(0)}$. 
The calculated results shown in Figs. 1-4 are common for the two integral paths (1) and (2).
%\newpage
 
 On the other hand, the convergence behaviors of the effective mass are different for the two integral paths.
\begin{figure}
\centerline{\includegraphics[width=8cm]{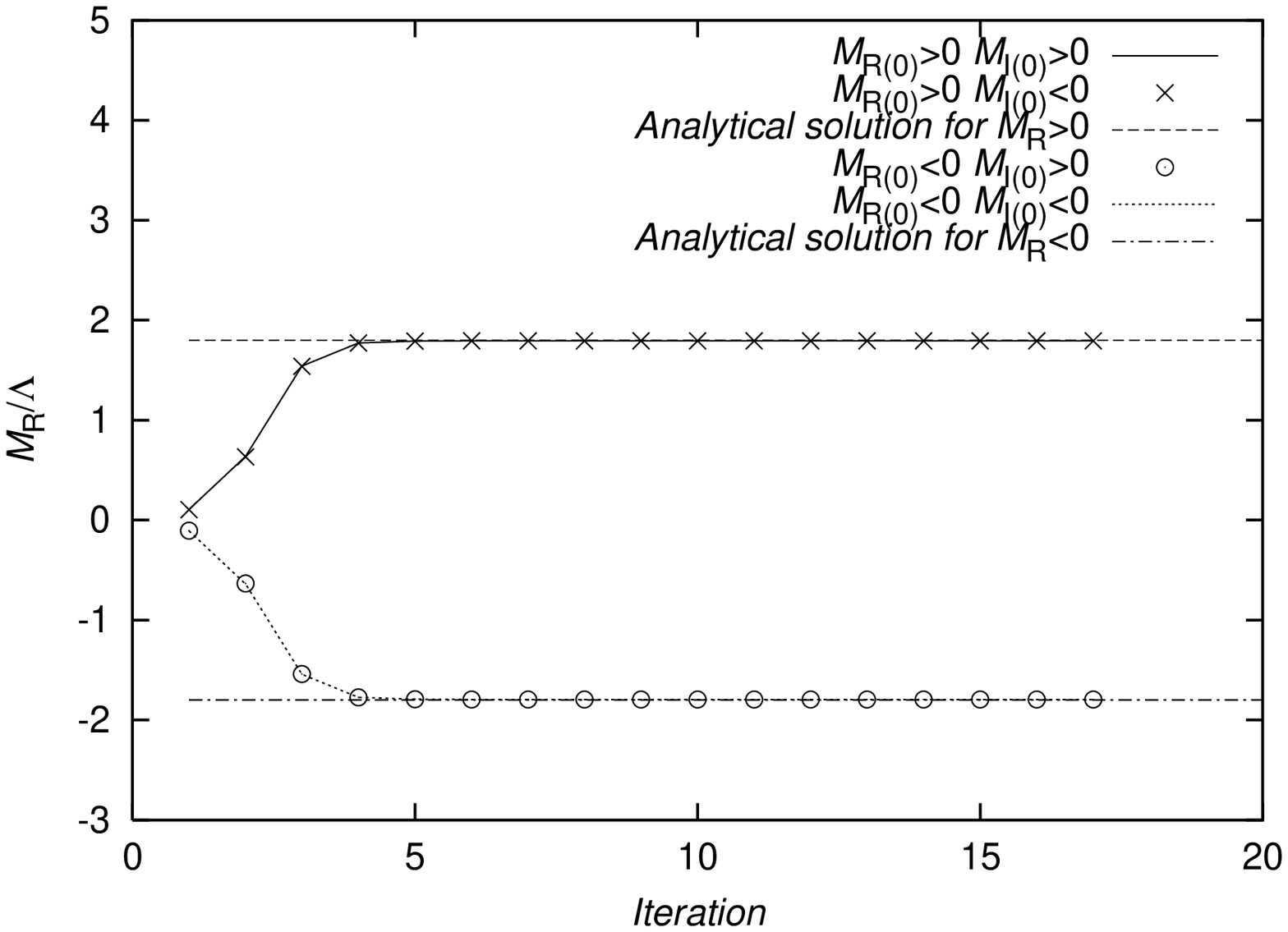}}
\caption{ The convergence  behaviors  of $M_{\rm R}/\Lambda$ for the integral path (1). The straight lines denote the analytical solutions of the real part of the effective mass divided by $\Lambda$. The horizontal axis denotes the number of iterations.}
%\label{Fig.1}
\end{figure} 
\begin{figure}
\centerline{\includegraphics[width=8cm]{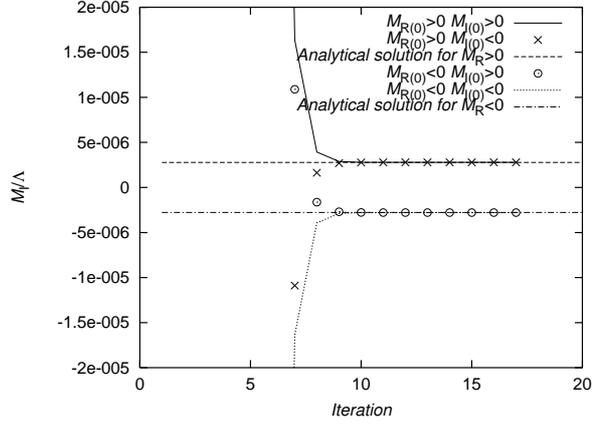}}
\caption{ The convergence  behaviors  of $ M_{\rm I}/\Lambda$ for the integral path (1). The straight lines denote the analytical solutions of the imaginary part of the effective mass divided by $\Lambda$. The horizontal axis denotes the number of iterations.}
%\label{Fig.1}
\end{figure} 
 \hspace{10cm}
 
 For the integral path (1), the real and imaginary parts of the effective mass calculated by the SDE are shown in Figs.5 and 6, respectively.
As shown in Fig. 5, the convergent solutions split into two values depending on the  respective signs of $M_{\rm R(0)}/\Lambda$.
%\newpage
As shown in Fig.6, the imaginary part of the effective mass is small and it depends on $\varepsilon$.  
Moreover, $ M_{\rm I}/\Lambda$ initially behaves according to the sign of each $M_{\rm I(0)} /\Lambda $, but the convergent solutions depend on the   respective signs of $M_{\rm R(0)} /\Lambda $.
\begin{figure}
\centerline{\includegraphics[width=8cm]{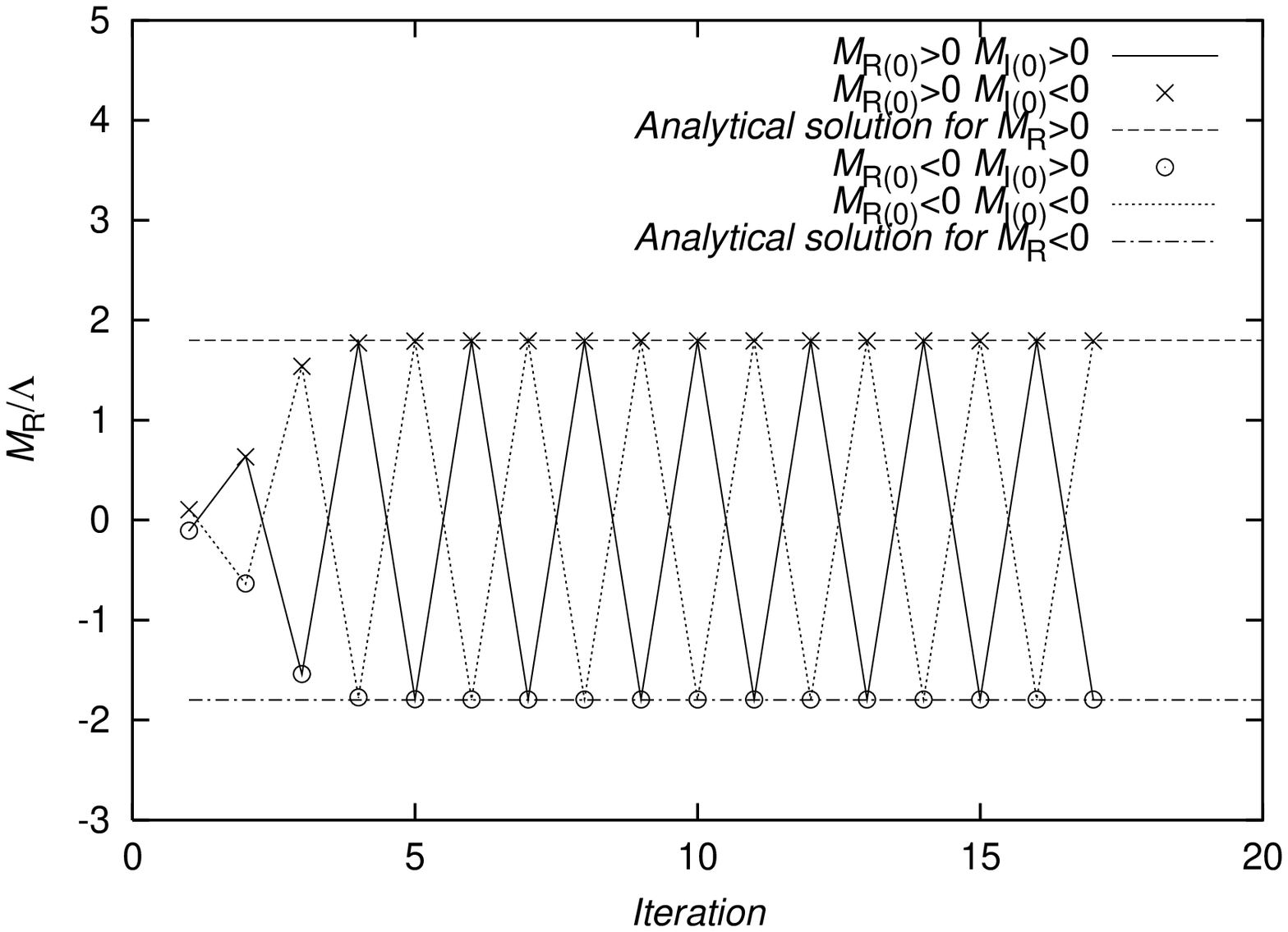}}
\caption{ The convergence   behaviors   of $M_{\rm R}/\Lambda$ for the integral path (2). The straight lines denote the analytical solutions of the real part of the effective mass divided by $\Lambda$. The horizontal axis denotes the number of iterations.}
%\label{Fig.1}
\end{figure} 
\begin{figure}
\centerline{\includegraphics[width=8cm]{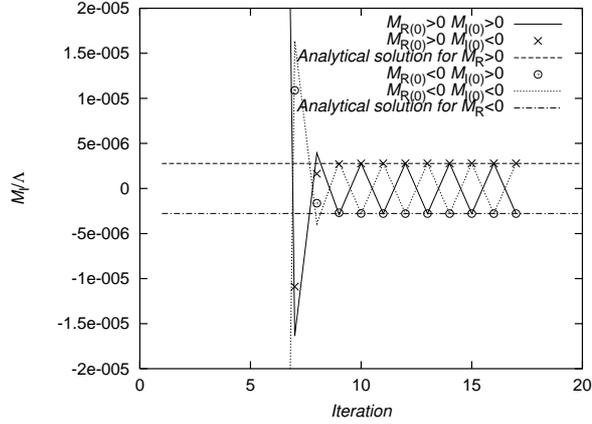}}
\caption{ The convergence   behaviors   of $ M_{\rm I}/\Lambda$ for the integral path (2). The straight lines denote the analytical solutions of the imaginary part of the effective mass divided by $\Lambda$. The horizontal axis denotes the number of iterations.}
%\label{Fig.1}
\end{figure} 
 \hspace{10cm}

   For the integral path (2), the convergence behaviors of the effective mass calculated by the SDE are shown in Figs.7 and 8, respectively.
As shown in Figs. 7 and 8, the convergent   solutions split   into two values depending on the   respective signs   of $M_{\rm R(0)}/\Lambda$ for $(M^2)_{\rm I(0)} <0$. However, for $(M^2)_{\rm I(0)}>0$, the iterated values are oscillated.

\section{The strong coupling QED with the IE approximation} 

\vspace{5mm}

\subsection{The SDE for the electron mass function}

\vspace{5mm}

In this section, we consider a slightly more non-trivial example, the strongly coupled QED.

We write the self-energy of the electron as
\begin{eqnarray}
\Sigma(P)=a(P)\slsh{P}+b(P)
\end{eqnarray}
with a   four-dimensional   momentum $P=(p_0,{\bf p})$, which is given by \begin{eqnarray}
 -i\Sigma(P)=(-ie)^2\int{d^4Q \over (2\pi)^4}\gamma^{\mu}iS(Q)\Gamma^{\nu}iD_{\mu\nu}(K),
\end{eqnarray}
  in one-loop approximation.   Here, $S(Q)$ and $D_{\mu\nu}(K)$ are propagators of a electron with $Q=(q_0,{\bf q})$ and   that of   a photon with $K= (k_0,{\bf k})= P-Q=(p_0-q_0,{\bf p-q})$, respectively

The effective propagator of the electron is given by
\begin{eqnarray}
iS(Q) = {i \over \slsh{Q}-\Sigma(Q)+i\varepsilon}={i \over (1-a(Q))\slsh{Q}-b(Q) +i\varepsilon }
\end{eqnarray}
with a massless electron.
 For simplicity, we take $\Sigma(P)=b(P)\equiv M(P)$ with $a(P)=0$ and the electron-photon  vertex with $\Gamma_{\mu}=\gamma_{\mu}$ in the Landau gauge.
Here, the photon propagator in the Landau gauge with quenched approximation is given as 
\begin{eqnarray}
iD_{\mu\nu}(K)=\left(-g_{\mu\nu}+{K_{\mu}K_{\nu} \over K^2}\right){i \over K^2+i\epsilon}.
\end{eqnarray}
 Using the approximation described above, the mass function is given by
\begin{eqnarray}
M(P)={1 \over 4}Tr[\Sigma(P)]
= ie^2 \int{d^4Q \over (2\pi)^4}{-3M(Q) \over  Q^2-M^2(Q)+i\varepsilon}{1 \over K^2+i\varepsilon}.
\end{eqnarray}

In the following calculations, we use the instantaneous exchange approximation (IEA) as $k_0=0$, in which the mass function does not depend on $p_0$ and $q_0$.

Integrating over the angle of the momentum ${\bf q}$, we obtain the SDE with the 
IEA as
\begin{eqnarray}
M(p)=i{3 \alpha \over 4\pi^2} \int_{\delta}^{\Lambda} dq{q \over p}\int dq_0 {M(q) \over  q_0^2-q^2-M^2(q)+i\varepsilon}\int_{\eta_-^2}^{\eta_+^2} {dk^2 \over k^2-i\varepsilon}
\end{eqnarray}
with $p=|{\bf p}|, q=|{\bf q}|, k=|{\bf k}|,\alpha=e^2/(4\pi)$ and $\eta^2_{\pm}=(p\pm q)^2$, respectively.
Using
\begin{eqnarray}
 I(q) \equiv i\int dq_0 {1 \over  q_0^2-q^2-M^2(q)+i\varepsilon}
=i\int dq_0 {1 \over  (q_0-E(q))(q_0+E(q))},
\end{eqnarray}
we can write the SDE as 
\begin{eqnarray}
M(p)= {3 \alpha \over 4\pi^2} \int_{\delta}^{\Lambda} dq{q \over p}M(q)I(q)\int_{\eta_-^2}^{\eta_+^2} {dk^2 \over k^2-i\varepsilon}.
\end{eqnarray}
Extending the $M(p),M(q)$ and $q_0$ to the complex values, we can evaluate $I(q)$ by the method in the previous section, which is given as
\begin{eqnarray}
I(q)=s(q)\pi{1 \over  E(q)}
\end{eqnarray}
with $s(q)=1$ for the integral path (1) and $s(q)=-E_{\rm I}(q)/| E_{\rm I}(q)|$ for the integral path (2), respectively.
Therefore, the SDE for the mass function is given by
 \begin{eqnarray}
M(p)={3 \alpha \over 4\pi} \int_{\delta}^{\Lambda} dq {q \over p} s(q)M(q) {1\over  E(q)}\int_{\eta_-^2}^{\eta_+^2} {dk^2 \over k^2-i\varepsilon}.
\end{eqnarray}
Integrating over $k^2$ in Eq.(3$\cdot$10), we have
\begin{eqnarray}
 J(p,q) \equiv \int_{\eta_-^2}^{\eta_+^2} {dk^2 \over k^2-i\varepsilon}
\equiv J_{\rm R}(p,q)+i J_{\rm I}(p,q)
\end{eqnarray}
with
\begin{eqnarray}
J_{\rm R}(p,q)={1 \over 2}\log{\eta_+^4+\varepsilon^2 \over \eta_-^4+\varepsilon^2}
\end{eqnarray}
and
\begin{eqnarray}
J_{\rm I}(p,q)=\arctan{\eta_+^2 \over \varepsilon}-\arctan{\eta_-^2 \over \varepsilon},
\end{eqnarray}
 respectively.

Therefore, the SDE for the mass function is given by 
\begin{eqnarray}
M(p)= {3 \alpha \over 4\pi} \int_{\delta}^{\Lambda} dq{q \over p}s(q){M(q) E^*(q)\over |E(q)|^2}J(p,q).
\end{eqnarray}
From the above formula,   the real and imaginary parts   of $M(p)$ are given by
$$ M_{\rm R}(p)= {3 \alpha \over 4\pi} \int_{\delta}^{\Lambda} dq{q \over p}s(q){[M(q)E^*(q) J(p,q)]_{\rm R}\over |E(q)|^2}$$
\begin{eqnarray}
= {3 \alpha \over 4\pi} \int_{\delta}^{\Lambda} dq{q \over p}s(q){ [M(q)E^*(q)]_{\rm R}J_{\rm R}(p,q)-[M(q)E^*(q)]_{\rm I}J_{\rm I}(p,q)\over |E(q)|^2}
\end{eqnarray}
and
$$M_{\rm I}(p)= {3 \alpha \over 4\pi} \int_{\delta}^{\Lambda} dq{q \over p}s(q){[M(q)E^*(q) J(p,q)]_{\rm I}\over |E(q)|^2}$$
\begin{eqnarray}
= {3 \alpha \over 4\pi} \int_{\delta}^{\Lambda} dq{q \over p}s(q){ [M(q)E^*(q)]_{\rm R}J_{\rm I}(p,q)+[M(q)E^*(q)]_{\rm I}J_{\rm R}(p,q)\over |E(q)|^2}
\end{eqnarray}
with
\begin{eqnarray} 
[M(q)E^*(q)]_{\rm R}=M_{\rm R}(q)E_{\rm R}(q)+M_{\rm I}(q)E_{\rm I}(q)
\end{eqnarray}
and
\begin{eqnarray} [M(q)E^*(q)]_{\rm I}=M_{\rm I}(q)E_{\rm R}(q)-M_{\rm R}(q)E_{\rm I}(q),
\end{eqnarray}
 respectively.

\subsection{Numerical results}

\vspace{5mm}
As can be seen in the previous example, it is possible to use the SDE for the integral path (1) to find the poles of the effective propagator.
Therefore, we present the results only for the integral path (1). 
Here, we show the convergence behaviors of the real and imaginary parts of the energy described by $E_{\rm R}(p) $ and $E_{\rm I}(p) $  with the momentum $p=\delta$, $\alpha=1,\delta/\Lambda=10^{-3}$ and $\varepsilon/\Lambda^2=10^{-4}$.\footnote{$E_{\rm R}(p) $ and $E_{\rm I}(p) $ are calculated using  $M_{\rm R}(p)$ and $M_{\rm I}(p)$ obtained by the SDE. } 
\newpage
In the following calculations, we set four initial   input   values for the mass functions as $ M_{\rm R(0)}= \pm 0.01\Lambda$ and $M_{\rm I(0)}=\pm 0.01\Lambda$, respectively. 
\begin{figure}
\centerline{\includegraphics[width=8cm]{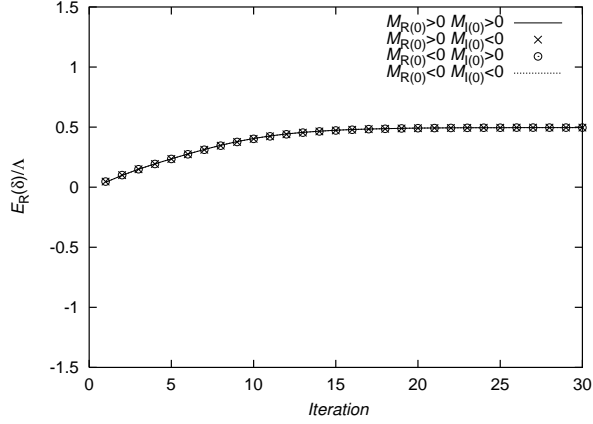}}
\caption{ The convergence    behaviors    of $ E_{\rm R}(p)/\Lambda$ with $p=\delta$. The horizontal axis denotes the number of iterations.}
%\label{Fig.1}
\end{figure}
%\hspace{28cm}
%\newpage
\begin{figure}
\centerline{\includegraphics[width=8cm]{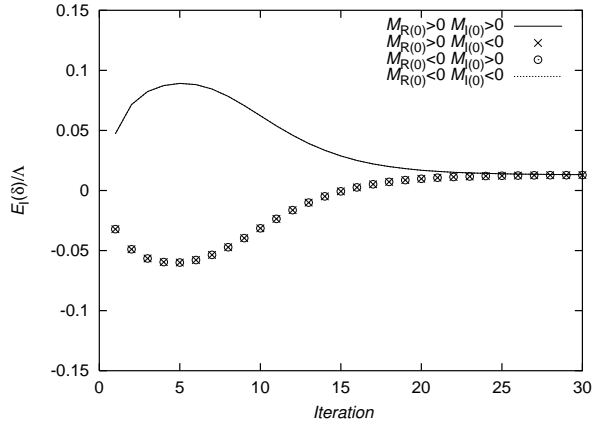}}
\caption{ The convergence    behaviors   of $E_{\rm I}(p)/\Lambda$ with $p=\delta$. The horizontal axis denotes the number of iterations.}
%\label{Fig.1}
\end{figure} 
\hspace{13cm}

In Fig.9, since the real part of the energy is defined to be positive, the numerical results do not depend on the   signs   of initial   input   values of the mass   function.  
In Fig.10, the imaginary part of the energy converges to $E_{\rm I}(\delta)>0$, in which the convergence   behaviors depend   on the respective signs of $(M^2)_{\rm I(0)}$. \footnote{In the numerical calculations, we rejected the integral point $p=q$, which corresponds to the infrared singular point for the photon propagator.}

\begin{figure}
\centerline{\includegraphics[width=8cm]{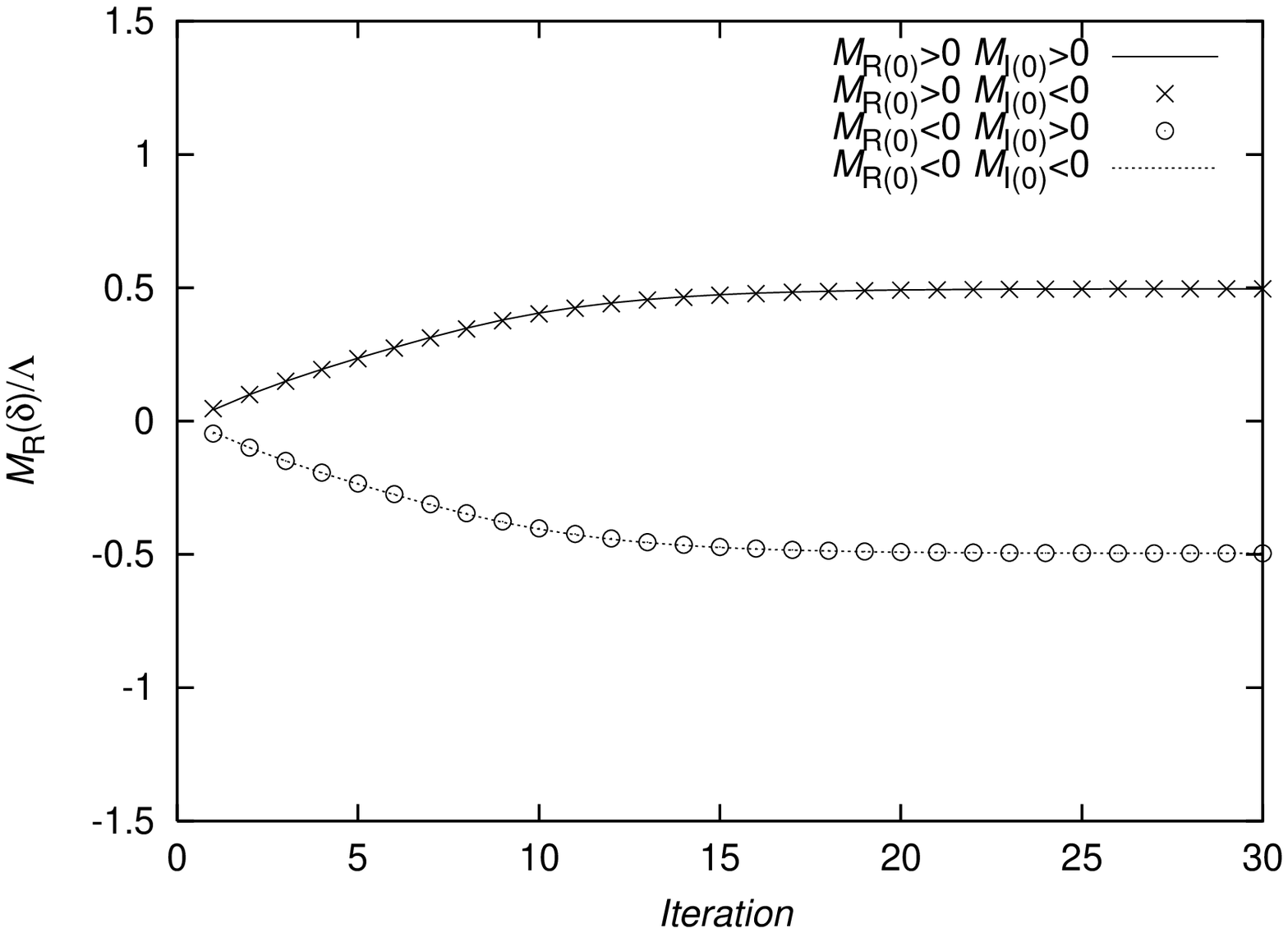}}
\caption{ The convergence   behaviors   of $ M_{\rm R}(p)/\Lambda$ with $p=\delta$. The horizontal axis denotes the number of iterations.}
%\label{Fig.1}
\end{figure}
%\hspace{28cm}
%\newpage
\begin{figure}
\centerline{\includegraphics[width=8cm]{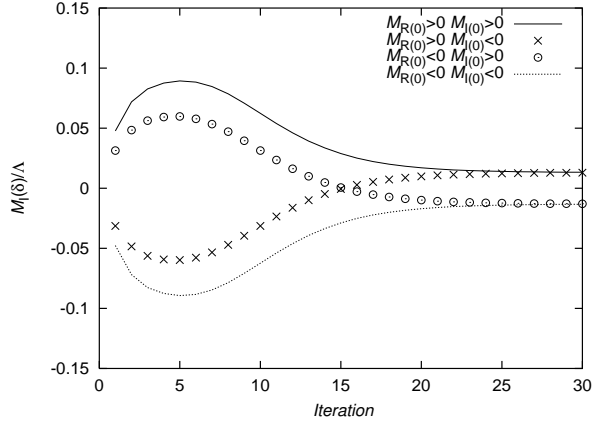}}
\caption{ The convergence   behaviors   of $M_{\rm I}(p)/\Lambda$ with $p=\delta$. The horizontal axis denotes the number of iterations.}
%\label{Fig.1}
\end{figure} 
\hspace{13cm}

The real and imaginary parts of the mass function calculated by the SDE with $p=\delta$ are shown in Figs.11 and 12, respectively.
As shown in Fig. 11, the convergent   solutions split   into two values depending on the   respective signs of $M_{\rm R(0)}/\Lambda$.

As shown in Fig.12, $ M_{\rm I}(\delta)/\Lambda$ initially behaves according to the sign of each $M_{\rm I(0)} /\Lambda $, but the convergent   solution depends   on the   respective signs of  $M_{\rm R(0)} /\Lambda $.
The   behaviors   shown above are similar to   those   for the first model in the previous section.

\begin{figure}
\centerline{\includegraphics[width=8cm]{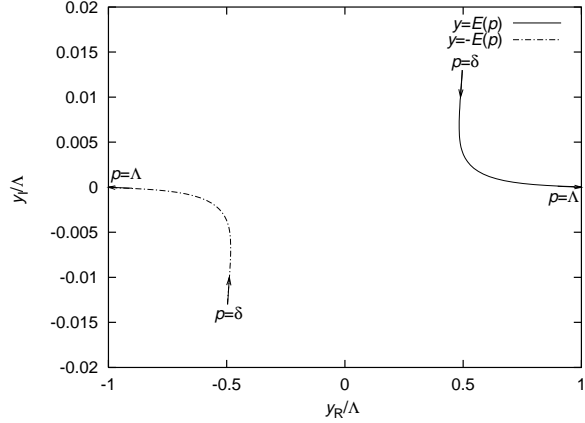}}
\caption{ The location of the poles of the effective propagator with $\alpha=1$ and $\varepsilon/\Lambda^2=10^{-4}$ from $p=\delta$ to $p=\Lambda$. }
%\label{Fig.1}
\end{figure} 
In Fig.13, we present the location of the poles of the effective propagator with $\alpha=1,\delta/\Lambda=10^{-3}$ and $\varepsilon/\Lambda^2=10^{-4}$ from $p=\delta$ to $p=\Lambda$, in which the denominator of the effective propagator is defined as
\begin{eqnarray} 
D(y,p)=y^2-p^2-M^2(p)+i\varepsilon=(y-E(p))(y+E(p))
 \end{eqnarray}
with $y=y_{\rm R}+iy_{\rm I}$. The arrows denote the momentum dependence of the poles of the effective propagator.

As shown in Fig.13, since $E_{\rm I}(p)>0$ for $\delta \leq p \leq \Lambda$, the solution obtained by the SDE in Eq.(3$\cdot$14) with $s(q)=-E_{\rm I}(q)/| E_{\rm I}(q)|$ for the integral path (2) may  oscillate. In Appendix B, we present an alternative method to obtain the mass function in Minkowski space, in which we use Eq.(3$\cdot$14) 
with $s(q)=-E_{\rm I}(q)/|E_{\rm I}(q)|$. However, in order to evaluate the right-hand side of Eq.(3$\cdot$14), we use the solution of the mass function $M_{(\rm E)}(p)$ obtained by the SDE for the integral path (1).
In our present case, the mass function $M_{(\rm M)}(p)$ in  Minkowski space is given as $M_{(\rm M)}(p)=- M_{(\rm E)}(p)$. Therefore, the relation between the energy  $E_{(\rm M)}(p)$ in Minkowski space and the energy $E_{(\rm E)}(p)$ in Euclid space is given as $E_{(\rm M)}(p)= E_{(\rm E)}(p)$.%\footnote{For $\varepsilon \rightarrow 0$, $E_{\rm I}(\delta)$ approaches to 0 from $E_{\rm I}(\delta)>0$.} 

\section{Summary and Comments}

In this paper, we investigated a method, in which the mass and energy of the Schwinger-Dyson equation (SDE) are extended to the complex plane, and the propagator in the integrand of the self-energy is integrated with two different integral paths in the complex energy plane.
 Then we examined the properties of the solutions obtained by the SDE in the complex plane.

We explained our formulation and analysis method using two simple examples.
First, in order to illustrate our method, we examined a simple SDE for the effective mass which does not depend on the energy and momentum.
We solved this model analytically and compared with the numerically determined solutions obtained by the SDE.
 Then we compered the effective mass and energy calculated in two different integral paths in the complex energy plane.
First, we investigated the effect of the momentum cutoff   for the effective mass.  Though the infrared cutoff $\delta$ on the momentum is an artificial parameter for numerical calculations, this example suggests a possibility of changing the critical point in physical systems with restricted low-momentum region. 
  In this model, the imaginary part of the energy is zero and the poles of the effective propagator are on the real axis, which is different situation in the strongly coupled quantum electrodynamics (QED) and the quantum chromodynamics (QCD) pointed out in Ref.[5-9], in which the poles are not on the real axis.

 We also investigated the dependence of the solutions obtained by the SDE on the initial input   values.  
The effective mass obtained by the SDE depends on the   respective signs of the initial input   values.   From our calculations, the SDE may lead to multiple solutions depending on the initial input values.
Moreover, for the integral path including the real axis, the sign of effective mass is not determined for some initial input values. 
These results suggest that the calculation of the SDE requires careful choice of the   initial   input values. 

As a second example, we investigated the strongly coupled QED with the instantaneous exchange approximation (IEA).  In this model, the mass function depends on the momentum.

  We found that the imaginary part of the energy $E_{\rm I}$ is positive value over the defined momentum range. Although $E_{\rm I}$ decreases as $E_{\rm I} \rightarrow +0$ for $E_{\rm R}>0$ with $\varepsilon \rightarrow +0$, the effective propagator has poles which are different from the pole locations in the  usual Feynman propagator, where $ E_{\rm I}=-\varepsilon \rightarrow -0$ for $E_{\rm R}>0$ with $\varepsilon \rightarrow +0$. The pole location of the Feynman propagator relates to the time-evolution of particle propagation. For example, a positive energy solution of particle propagates forward in time. In our case, the relation with the time-ordered product of the wave-function in the definition of the two-point Green's function of the Feynman propagator appears to be broken. 
If we take the Fourier transform of our modified propagator from momentum space to space-time, for the positive (negative) energy solution, we integrate over the upper (lower) half of the complex energy plane converging for time $t<0~ (t>0)$.
Therefore, we have a propagator for positive (negative) energy solution which is defined in $t<0~ (t>0)$. 
From our example, it may be expected that the effective propagator behaves anomalously in non-perturbative region.

In order to get more practical results, we need to investigate the mass function, which depends on   the energy component of the four-dimensional momentum.   
The way to do this is to calculate the self-energy without the IEA.  

Although our models   studied   in this paper may be a little too simple for real physical situations, we expect that our method and analyses will give some hints when applying the SDE to other models such as QCD in non-perturbative region.

%\section*{Acknowledgements}

\vspace{5mm}
\begin{center}
{\large \bf Appendix A. Complex mass and energy}
\end{center}

\vspace{5mm}

In order to solve the SDE in complex energy plane, we need the explicit forms of the complex mass and energy.

We define the complex mass as $M=M_{\rm R}+iM_{\rm I}$ and  the squared of the mass as $M^2=(M^2)_{\rm R}+i(M^2)_{\rm I}$. Here, $(M^2)_{\rm R}$ and $(M^2)_{\rm I}$ are given by
$$(M^2)_{\rm R}=M_{\rm R}^2-M_{\rm I}^2,~~(M^2)_{\rm I}=2M_{\rm R}M_{\rm I}.$$
The squared energy $E^2$ is defined by 
$$E^2=q^2+M^2-i\varepsilon \equiv (E^2)_{\rm R}+i(E^2)_{\rm I}.$$
with
$$ (E^2)_{\rm R}=q^2+(M^2)_{\rm R},~~(E^2)_{\rm I}=(M^2)_{\rm I}-\varepsilon.$$
On the other hand, using the complex energy  $E=E_{\rm R}+iE_{\rm I}$, $(E^2)_{\rm R}$ and $(E^2)_{\rm I}$ are written as 
$$ (E^2)_{\rm R}=E_{\rm R}^2-E_{\rm I}^2,~~(E^2)_{\rm I}=2E_{\rm R}E_{\rm I}.$$
Therefore the imaginary part of the energy is given by
$$ E_{\rm I}={(E^2)_{\rm I} \over 2E_{\rm R}}.$$
Substituting above equation to $(E^2)_{\rm R}=E_{\rm R}^2-E_{\rm I}^2$,
we have a quadratic equation for $ E_{\rm R}^2$ as
$$ (E_{\rm R}^2)^2-E_{\rm R}^2(E^2)_{\rm R}-(E^2)_{\rm I}^2/4=0. $$
The solution of the equation for $E_{\rm R}^2 > 0$ is given by 
$$ E_{\rm R}^2={(E^2)_{\rm R}+ |E^2| \over 2}$$
with $|E^2|=\sqrt{[(E^2)_{\rm R}]^2+[(E^2)_{\rm I}]^2}$.

Therefore, we get $ E_{\rm R}$ and $ E_{\rm I}$ as 
$$ E_{\rm R}=\sqrt{(E^2)_{\rm R}+|E^2| \over 2},~~E_{\rm I}={(E^2)_{\rm I} \over 2E_{\rm R}}$$
for $E_{\rm R} > 0$.

\vspace{5mm}
\begin{center}
{\large \bf Appendix B. Alternative method to evaluation the mass function in Minkowski space}
\end{center}

\vspace{5mm}

Here, the SDE with IEA in the complex plane is given as
$$M(p)=i{3 \alpha \over 4\pi^2} \int dq{q \over p}\oint dz {M(q) \over  z^2-q^2-M^2(q)+i\varepsilon}\int_{\eta_-^2}^{\eta_+^2} {dk^2 \over k^2-i\varepsilon}$$
$$ \equiv {3 \alpha \over 4\pi^2} \int dq{q \over p}M(q)I(q)\int_{\eta_-^2}^{\eta_+^2} {dk^2 \over k^2-i\varepsilon}$$
with 
$$ I(q)= i\oint dz {1 \over  z^2-q^2-M^2(q)+i\varepsilon}
=i\oint dz {1 \over  (z-E(q))(z+E(q))},$$
 where $M(p),M(q),E(q)$ and $z$ are the complex values.
Defining 
$$ I_{(\rm M)}(q) \equiv i\int^{\infty}_{-\infty} dq_0 {1 \over  q_0^2-q^2-M^2(q)+i\varepsilon},$$
$$ I_{(\rm E)}(q) \equiv i\int_{\infty}^{-\infty} idq_4 {1 \over  -q_4^2-q^2-M^2(q)+i\varepsilon}
=\int_{\infty}^{-\infty} dq_4 {1 \over  q_4^2+q^2+M^2(q)-i\varepsilon},$$
$$ I_{(\rm C_+)}(q) \equiv \lim_{\Lambda_0\rightarrow \infty}i\int_0^{\pi/2}  id\theta{\Lambda_0 e^{i\theta} \over  \left(\Lambda_0 e^{i\theta}\right)^2-q^2-M^2(q)+i\varepsilon}$$
 and 
$$ I_{(\rm C_-)}(q) \equiv \lim_{\Lambda_0\rightarrow \infty}i\int^{\pi}_{3\pi/2}  id\theta{\Lambda_0 e^{i\theta} \over  \left(\Lambda_0 e^{i\theta}\right)^2-q^2-M^2(q)+i\varepsilon},$$
 we chose the following integral path as
$$I(q)=I_{(\rm M)}(q) +I_{(\rm C_+)}(q)+I_{(\rm E)}(q)+I_{(\rm C_-)}(q)=R_{+}(q)+R_{-}(q),$$
where $R_{\rm \pm}(q)$ are residues of the poles enclosed by the integral path,   which are given by   
$$ R_{+}(q)=i {2i\pi \over 2E(q)}\Theta(E_{\rm I}(q))=-{\pi \over E(q)}\Theta(E_{\rm I}(q))$$
 and
$$R_{-}(q)=i {-2i\pi \over -2E(q)}\Theta(E_{\rm I}(q))=-{\pi \over E(q)}\Theta(E_{\rm I}(q)),$$
respectively. Here, $\Theta(E_{\rm I}(q))$ is a step function for the imaginary part of the energy.
Since $ I_{(\rm C_{\pm})}(q)\rightarrow 0$ for $\Lambda_0\rightarrow \infty$, we have the following relation between the integration of the propagator by the Minkowski momentum $I_{(\rm M)}(q)$ and that by the Euclidean  momentum $I_{(\rm E)}(q)$ as
$$I_{(\rm M)}(q)=-I_{(\rm E)}(q)+R_{+}(q)+R_{-}(q),$$
 which gives
$$i\int^{\infty}_{-\infty} dq_0 {1 \over  q_0^2-q^2-M^2(q)+i\varepsilon}
=\int^{\infty}_{-\infty} dq_4 {1 \over  q_4^2+q^2+M^2(q)-i\varepsilon}
-{2\pi \over E(q)}\Theta(E_{\rm I}(q)).$$
Defining 
$$ M_{\rm (M)}(p)\equiv {3 \alpha \over 4\pi^2} \int dq{q \over p}M_{\rm (M)}(q)I_{\rm (M)}(q)\int_{\eta_-^2}^{\eta_+^2} {dk^2 \over k^2-i\varepsilon}$$
and
$$ M_{\rm (E)}(p)\equiv -{3 \alpha \over 4\pi^2} \int dq{q \over p}M_{\rm (E)}(q)I_{\rm (E)}(q)\int_{\eta_-^2}^{\eta_+^2} {dk^2 \over k^2-i\varepsilon},$$
 we have
$$M_{\rm (M)}(p)=M_{\rm (E)}(p)
-{3 \alpha \over 4\pi} \int dq{q \over p}M_{\rm (E)}(q)\left({2 \over E_{\rm (E)} (q)}\Theta((E_{\rm (E) } (q))_{\rm I})\right)\int_{\eta_-^2}^{\eta_+^2} {dk^2 \over k^2-i\varepsilon}.$$
Using the relation 
$$ \Theta((E_{\rm (E) } (q))_{\rm I})={1 \over 2}\left(1+{(E_{\rm (E) } (q))_{\rm I} \over |(E_{\rm (E) } (q))_{\rm I}|}\right),$$
 we obtain 
$$M_{\rm (M)}(p)={3 \alpha \over 4\pi} \int dq{q \over p}s_{\rm (E)}(q)M_{\rm (E)}(q) {1\over  E_{\rm (E)} (q)}\int_{\eta_-^2}^{\eta_+^2} {dk^2 \over k^2-i\varepsilon}$$
 with 
$$ s_{\rm (E)}(q)=- {(E_{\rm (E) } (q))_{\rm I} \over |(E_{\rm (E) } (q))_{\rm I}|}.$$
Here, the solution of $M_{\rm (E)}(q)$ is obtained by the SDE for the integral path (1) as
$$ M_{\rm (E)}(p) ={3 \alpha \over 4\pi} \int dq{q \over p}M_{\rm (E)}(q){1\over  E_{\rm (E)} (q)}\int_{\eta_-^2}^{\eta_+^2} {dk^2 \over k^2-i\varepsilon},$$
then the energy $E_{\rm (E)} (q) $ is evaluated with $ M_{\rm (E)}(q)$.
 For example, if $(E_{\rm (E)} (q))_{\rm I}<0 $ for all $q$, we have $ M_{\rm (M)}(p) = M_{\rm (E)}(p) $. In this case, there are no poles inside the integral path for all $q$. Then, the Wick rotation from the real axis to the imaginary axis is valid. Conversely, if $(E_{\rm (E)} (q))_{\rm I}>0 $ for all $q$, then $ M_{\rm (M)}(p) = -M_{\rm (E)}(p) $. In this case, there are poles inside the integral path for all $q$.

\vspace{5mm}

%-------------------------------------------------------------%                 %  REFERENCES
%%---------------------------------------------------------------------------
%\newpage
%\begin{thebibliography}{99}
\begin{flushleft} 
{\bf References }
\end{flushleft}

\vspace{2mm}

\begin{description}
\item{[1]}  F.J.Dyson, Phys.Rev.{\bf 75} (1949) ,1736.
\item{[2]}  J.S.Schwinger,Proc.Nat.Acad.Sci.{\bf 37} (1951),452.
\item{[3]} Y. Hayashi and K.-I. Kondo, Phys. Rev. {\bf D103}, L111504 (2021), [arXiv:2103.14322].
\item{[4]} Y. Hayashi and K.-I. Kondo, Phys. Rev.{\bf D104},074024(2021), [arXiv:2105.07487].
\item{[5]}  P. Maris, Phys.Rev.{\bf D50},4189(1994).
\item{[6]}  P. Maris, Phys.Rev.{\bf D52},6087(1995) [arXiv:hep-ph/9508323].
\item{[7]}  P. Maris and H.A.Holties, Int.J.Mod.Phys.{\bf A7} (1992),5369.
\item{[8]}  S.Strauss, C.S.Fischer and C.Kellermann, Phys. Rev. Lett. {\bf 109} (2012),252001 [arXiv:1208:6239 [hep-ph]].
\item{[9]} C.S.Fischer and M.Q.Huber, Phys. Rev.  {\bf D102} (2020),094005 [arXiv:2007.11505].
\item{[10]} N.Dorey and N.E.Mavromatos, Phys.Lett.{\bf B266}, 163(1991).
\item{[11]} K.Fukazawa, T.Inagaki, S.Mukaigawa and T.Muta, Prog. Theor. Phys. {\bf 105},979(2001) [arXiv:hep-ph/9910305]. 
\item{[12]} D.J.Gross and A.Neveu, Phys. Rev.  {\bf D10} (1974),3235.
\end{description}

\end{document}